# High-low level support vector regression prediction approach (HL-SVR) for data modeling with input parameters of unequal sample sizes

Maolin Shi[a], Wei Sun[a], Xueguan Song[a]*, Hongyou Li[b]

[a] School of Mechanical Engineering, Dalian University of Technology, Linggong Road, Dalian, China, 116024

[b] College of Mechanical Engineering and Automation, Huaqiao University, Jimei Avenue, Xiamen, China, 361021

**Abstract**

Support vector regression (SVR) has been widely used to reduce the high computational cost of computer simulation. SVR assumes the input parameters have equal sample sizes, but unequal sample sizes are often encountered in engineering practices. To solve this issue, a new prediction approach based on SVR, namely as high-low-level SVR approach (HL-SVR) is proposed for data modeling of input parameters of unequal sample sizes in this paper. The proposed approach is consisted of low-level SVR models for the input parameters of larger sample sizes and high-level SVR model for the input parameters of smaller sample sizes. For each training point of the input parameters of smaller sample sizes, one low-level SVR model is built based on its corresponding input parameters of larger sample sizes and their responses of interest. The high-level SVR model is built based on the obtained responses from the low-level SVR models and the input parameters of smaller sample sizes. Several numerical examples are used to validate the performance of HL-SVR. The experimental results indicate that HL-SVR can produce more accurate prediction results than conventional SVR. The proposed approach is applied on the stress analysis of dental implant, which the structural parameters have massive samples but the material of implant can only be selected from several Ti and its alloys. The prediction performance of the proposed approach is much better than the conventional SVR. The proposed approach can be used for the design, optimization and analysis of engineering systems with input parameters of unequal sample sizes.

## 1. Introduction

Computer simulation techniques have been widely used in the design, analysis and optimization of engineering systems, since its ability of presenting the true physics of phenomena [1-3]. Recently, the need for high-fidelity computer simulations has been growing greatly in various engineering applications due to their high-level of accuracy, but the computational cost is increasing dramatically as well [4]. To address the challenge of reducing the computational cost of computer simulation and the need of modeling the engineering data, a statistical learning technique, support vector regression (SVR), are widely used to represent computationally expensive computer simulation [5-6]. In recent years, many support vector regression-based prediction approaches have been proposed and successfully applied in different areas of engineering [7-12]. Eisenhower used support vector regression to predict the building energy consumption. The results indicate that the support vector regression can provide accurate prediction results and saves greatly computing time compared with the conventional building energy model [7]. Pan utilized support vector regression for vehicle lightweight design and demonstrated that support vector regression is available for function approximation of highly nonlinear crash problems [8]. Gryllias proposed a hybrid two stage one-against-all support vector regression approach for the automated diagnosis of defective rolling element bearings [9]. Andrés used support vector regression to replace the expensive computational fluid dynamics simulation and combine it with an evolutionary algorithm to optimize aeronautical wing profiles [10]. Samui applied least square support vector regression for the safety factor prediction of the slope, and demonstrated that least square support vector regression is a more robust model for slope stability analysis compared with artificial neural network [11]. Zheng proposed a modified support vector regression for the variable-fidelity data to replace the computationally expensive computer simulation [12]. Ghosh proposed a support vector regression-based metamodeling approach and applied it for efficient seismic reliability analysis of structure [13]. Nik used support vector regression to replace the finite element method to estimate the mechanical properties of composite laminates, and used it in the design optimization of composite laminates with variable stiffness [14]. Zhu utilized support vector regression to describe the nonlinear relationship between the crashworthiness of the vehicle and its structural parameters, and the experimental results indicate that support vector regression is a promising alternative for approximating highly nonlinear crash problems [15].

The above-mentioned works highlights the benefits of support vector regression for the design and optimization of engineering systems. However, most of the support vector regression-based prediction approaches assume that all the input parameters in the computer simulation have same sample sizes. But unequal sample sizes are widely present among input parameters in engineering practices, in other words, some input parameters are with smaller sample sizes but some input parameters have massive samples. For example, the stress at implant-bone interface of dental implant is affected by the structure and material of implant [16]. The structure of dental implant can be arbitrarily changed which means that the structural parameters can obtain massive samples, but the samples of material parameters are much smaller since the implant material can be only selected from a small amount of Ti or its alloys [17-18]. Thus, the sample sizes of material parameters are much smaller than those of structural parameters for dental implant. To solve this issue that building support vector regression for the parameters of unequal sample size, some researchers consider the input parameters of smaller sample sizes as qualitative variables, and the input parameters of larger sample sizes are considered as quantitative variables. The study focus is the ways to construct the correlation matrix of qualitative variables and to combine it with that of quantitative variables. McMillian, Joseph and Delaney used restrictive correlation function to construct the correlation function of qualitative variables [19-20]. Such an approach can simplify the computational complexity for the model estimation, but the restrictive correlation functions lack flexibility to quantify general correlation structure of qualitative variables. To solve this problem, Qian developed a general framework of constructing an unrestrictive correlation structure for qualitative variables [21-22]. However, they consider the input parameters of smaller sample size as qualitative variables, which means that they cannot handle the problem that when the input parameters of smaller sample sizes have a new value different from that of the training points. For example, one input parameter of smaller sample size in Ref [23] has three values -50, 0, 50 in the training points, thus the input parameter of smaller sample size has three quantitative levels as one, two and three, respectively. If the input parameter of smaller sample size of the new point has another value of 25, the above approaches cannot handle it. To solve this problem, a new support vector regression-based prediction approach, namely as high-low-level support vector prediction approach (HL-SVR), is proposed in this paper. The proposed approach is consisted of low-level SVR models for the input parameters of larger sample sizes and a high-level SVR model for the input parameters of smaller sample sizes. In this approach, we assume that the input parameters of smaller sample sizes have significant influence on the relationship between the response of interest and the input parameters of larger sample sizes. The above assumption can be easily understood in engineering practices. For instance, the selection of Ti alloy has considerable effect on the relationships between structural parameters and stress at implant-bone as discussed in the previous studies [16, 24]. The SVR models based on the input parameters of larger sample sizes and the corresponding responses of interest at each training point of the input parameters of smaller sample sizes can more accurately obtain their relationship. Thus, a new prediction strategy is proposed in this paper as follows. For each training point of the input parameters of smaller sample sizes, its corresponding input parameters of larger sample sizes and their responses are used to build a low-level SVR model. After that, the high-level SVR model is built based on the obtained responses from the low-level SVR models and the training samples of the input parameters of smaller sample sizes. The input parameters of larger sample sizes of the new point are first inputted into the low-level SVR models. Then, the input parameters of smaller sample sizes of the new point are inputted into the high-level SVR model, and the response of the new point is obtained. Thus, the HL-GPR approach can realize the accurate prediction for the input parameters of unequal sample sizes theoretically but not need to consider the input parameters of smaller sample sizes as quantitative factors.

The remaining of this paper is organized as follows. Section 2 introduces the details of support vector regression and the proposed approach. Section 3 presents the performance comparison between the proposed approach and conventional SVR on several numerical examples. In Section 4, a computer simulation case (stress analysis of dental implant) is used to validate the proposed approach. The last section concludes this work.

## 2. Proposed approach (HL-SVR)

2.1 Support vector regression

SVR is based on support vector machine (SVM) whose purpose is to evaluate the complex relationship between the input and the response of interest through mapping the data into a high-dimensional feature space. Let the *i*-th input be denoted by a dimensional vector, $\mathbf{x}_i = (x_{i1}, \ldots, x_{id})$, and its response, $y_i$, respectively. The regression model of SVR can be constituted as follows:

$$y = \omega^T \cdot \varphi(x) + b \tag{1}$$

where $\varphi$ denotes the feature map, $\omega$ is the weight vector and $b$ is the bias term. In SVR, it is necessary to minimize a cost function (*C*) containing a penalized regression error as shown below:

$$C = \frac{1}{2}\omega^T \cdot \omega + \frac{1}{2}\gamma \sum_{i=1}^{n} e_i^2 \tag{2}$$

The first part of cost function (2) is a weight decay which is used to regularize weight sizes and penalize large weights. The second part is the regression error for all training data. The parameter $\gamma$ determines the relative weight of this part as compared to the first part. To optimize the cost function (2), Lagrange multipliers methods is used as follows:

$$L(\omega, b, e: \alpha) = \frac{1}{2}\|\omega\|^2 + \gamma \sum_{i=1}^{n} e_i^2 - \sum_{i=1}^{n} \alpha_i \{\omega^T \cdot \varphi(x_i) + b + e_i - y_i\} \tag{3}$$

where $\alpha_i$ are Lagrange multipliers. Through setting the partial first derivatives to zero, the optimum solution can be obtained.

$$\frac{\partial L}{\partial \omega} = 0 \rightarrow \omega = \sum_{i=1}^{n} \alpha_i \varphi(x_i)$$

$$\frac{\partial L}{\partial b} = 0 \rightarrow \sum_{i=1}^{n} \alpha_i = 0$$

$$\frac{\partial L}{\partial e_i} = 0 \rightarrow \alpha_i = \gamma e_i, i = 1,2,\ldots,n$$

$$\frac{\partial L}{\partial \alpha_i} = 0 \rightarrow \omega^T \cdot \varphi(x_i) + b + e_i - y_i = \gamma e_i, i = 1,2,\ldots,n$$

Thus,

$$\omega = \sum_{i=1}^{n} \alpha_i \varphi(x_i) = \sum_{i=1}^{n} \gamma e_i \varphi(x_i) \tag{5}$$

where a positive definite kernel is used as follows:

$$K(x_i, x_j) = \varphi(x_i)^T \varphi(x_i) \tag{6}$$

The original regression model in (1) can be modified as follows:

$$y = \sum_{i=1}^{n} \alpha_i \, \varphi(x_i)^T \varphi(x) + b = \sum_{i=1}^{n} \alpha_i \, \langle \varphi(x_i)^T, \varphi(x) \rangle + b \tag{8}$$

For a point of $y_j$ to be evaluated it is:

$$y_j = \sum_{i=1}^{n} \alpha_i \, \langle \varphi(x_i)^T, \varphi(x_j) \rangle + b \tag{9}$$

The $\boldsymbol{\alpha}$ vector can be obtained from solving a set of linear equations:

$$\begin{bmatrix} K + \frac{1}{\gamma} & 1_N \\ 1_N^T & 0 \end{bmatrix} \begin{bmatrix} \alpha \\ b \end{bmatrix} = \begin{bmatrix} y \\ 0 \end{bmatrix} \tag{10}$$

And the solution is:

$$\begin{bmatrix} \alpha \\ b \end{bmatrix} = \begin{bmatrix} K + \frac{1}{\gamma} & 1_N \\ 1_N^T & 0 \end{bmatrix}^{-1} \begin{bmatrix} y \\ 0 \end{bmatrix} \tag{11}$$

In this paper, the radial basis function is used as the kernel function

$$e^{-\theta \|x_i - x_j\|^2} \tag{12}$$

## 2.2 HL-SVR

In this paper, a new prediction approach based on support vector regression is proposed for data modeling with input parameters of unequal sample sizes. We assume that the input parameters of smaller sample sizes ($Input_s$s) can represent the main characteristics of computer simulation and have significant influence on the relationship between the response of interest and the input parameters of smaller sample sizes ($Input_l$s). The above assumption can be easily understood in engineering practices. For instance, the selection of Ti alloy can affect the relationships between structure parameters and stress at implant-bone greatly as discussed in the previous studies. Since the relationship between $Input_l$s and response of interest varies greatly for different $Input_s$s, it is difficult using only one model to simulate these relationships simultaneously. The model based on the $Input_l$s and the corresponding responses of interest at each training point of $Input_s$s can more accurately obtain their relationship. Thus, a new prediction strategy is proposed as follows. For each training point of $Input_s$s, its training samples of $Input_l$s and their responses are used to build an SVR model, then the $Input_l$ of the new point is inputted into these SVR models to get the corresponding responses. After that, another SVR model is built based on the obtained responses and the training samples of $Input_s$s. Finally, the $Input_s$ of the new point is inputted into this model, and the output of the new point is obtained.

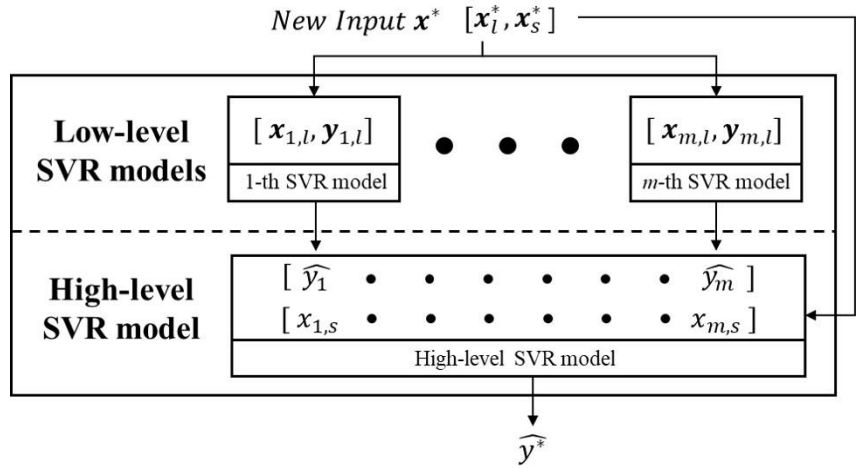

Figure 1 Framework of HL-SVR

Based on the proposed strategy above, a new high-low-level prediction approach based on support vector regression, namely as HL-SVR, is proposed as shown in Fig. 1. The main framework of this approach is constructed of two-level SVR models (high-level and low-level). In the low-level, $m$ SVR models are built for each training point of $Input_s$s based on its corresponding $Input_l^{Train}$s and the real responses $\overrightarrow{y^{train}}$, where $m$ is the number of $Input_s$s samples. To obtain the response of a new point ($x^*$), its $Input_l^{x^*}$ is first inputted into these $m$ SVR models and get the estimated responses ($\widehat{y_1}, \ldots, \widehat{y_m}$). Then its $Input_s^{x^*}$ is inputted into the high-level SVR model which is built based on ($\widehat{y_1}, \ldots, \widehat{y_m}$) and the training samples of $Input_s$s ($Input_{1s}^{Train}, \ldots, Input_{ms}^{Train}$), and the output $\widehat{y^*}$ is obtained.

## 3. Numerical examples

In this section, the performance of HL-SVR is validated and compared with SVR on several numerical examples. The benchmark functions of these examples are widely used to validate the performances of surrogate models. Design of Experiments (DoE) is used to generate sampling or training samples. Among many available DoE methods, the Latin Hypercube Sampling (LHS) has been proved capable of balancing the trade-off between accuracy and robustness by generating a near-random set of sample points [25]. In this paper, the ***R*** package *lhs* is used, and the training samples are generated as follows. The input parameters of each example are divided into two parts, input parameters of smaller sample size ($Input_s$s) and the input parameters of larger sample size ($Input_l$s). The training samples of $Input_s s$ are generated using *lhs*, whose size is 2~3*n*. Then, the training samples of $Input_l s$ are generated repetitively for each sample of $Input_s s$, whose size is 10n. Take numerical example 1 for instance, the $Input_s s$ are $x_{s,1}$ and $x_{s,2}$, and four training samples are generated using LHS which are [0.852, -0.153], [-0.376, 0.431], [0.277, 0.853] and [-0.523, -0.989]. After that, 10 training samples are generated four times for [0.852, -0.153], [-0.376, 0.431], [0.277, 0.853] and [-0.523, -0.989], respectively. The testing samples are generated similarly, but the sizes of $Input_s$ and $Input_l$ are 10n and 30n, respectively.

To compare the performance of the proposed approach and the SVR, Root mean squared error is used as follows:

$$RMSE = \sqrt{\frac{\sum_{i=1}^{n}(y_i - \hat{y}_i)}{n}} \qquad (13)$$

where *n* is the number of newly created validation samples, $y_i$ is the true results on the validation samples, and $\hat{y}_i$ is the corresponding approximate results. The smaller the value of *RMSE*, the better the prediction accuracy. For each numerical test problems, 30 times experiments are conducted in which the training and testing samples are generated repeatedly for each experiment.

*Example* 1. This example considers an experiment with two $Input_s s$ ($x_{s,1}$, $x_{s,2}$) and two $Input_l s$ ($x_{l,1}$, $x_{l,2}$), taking values on [-1,1]. The response of the experiments takes the following form:

$$y = (x_{l,1} - \frac{5.1(x_{s,1})^2}{4\pi^2} + \frac{5x_{l,2}}{\pi} - 6)^2 + 10\left(1 - \frac{1}{8\pi}\right)cos(x_{s,2}) + 10 \qquad (14)$$

The training samples of $Input_s s$ are generated using LHS, whose size is 2*n* for this function. Then, the training samples of $Input_l s$ are generated repetitively for each sample of $Input_s s$, whose size is 10*n*, respectively. The testing samples are generated similarly, but the sizes of $Input_s$ and $Input_l$ are 10n and 30n, respectively. The root means squared errors (RMSEs) of the testing samples are calculated for SVR and the proposed approach to assess their prediction accuracy. This procedure of data generation, model fitting, and assessment of prediction accuracy was repeated 30 times. Figure 2 compares the average RMSEs of 30 experiments. From this figure, it is observed that average RMSE of HL-SVR is smaller than that of SVR, which demonstrates that HL-SVR is able to provide more accuracy prediction results than SVR. In addition, the standard deviation (Std.) is as well reduced greatly by HL-SVR. The proposed approach outperforms SVR in terms of both accuracy and robustness.

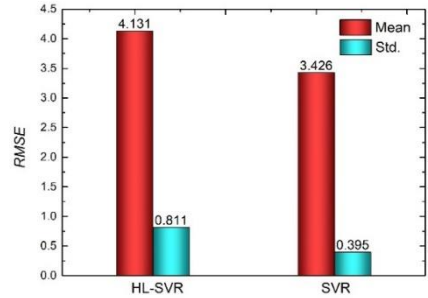

Figure 2 RMSE comparison between HL-SVR and SVR

*Example* 2. Consider another numerical example with one $Input_s$s and two $Input_l$s. The training data generated form the following function,

$$y = x_{l,1}^2\left(4 - 2.1x_{s,1}^2 + \frac{x_{s,1}^4}{3}\right) + x_{l,1}x_{s,1} + (4x_{l,1}^2 - 4)x_{l,2}^2 \tag{15}$$

where $0 \leq x_{s,1}, x_{l,1}, x_{l,2} \leq 1$. In each experiment, three training samples of $Input_s$s are generated using LHS, and then 20 training samples of $Input_l$s are generated repetitively for each sample of $Input_s$s. The testing sample sizes of $Input_s$ and $Input_l$ are 10 and 60, respectively. The RMSEs of the 30 times experiments are shown Fig. 3. From this figure, it is clearly seen that HL-SVR outperforms SVR since its corresponding RMSE is much lower than that of SVR. The HL-SVR prediction approach shows better prediction accuracy and robustness in this numerical example.

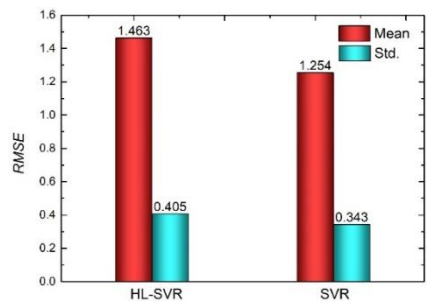

Figure 3 RMSE comparison between HL-SVR and SVR

*Example* 3. This example considers an experiment with four $Input_s$s and four $Input_l$s, taking values on [0, 1]. The response of the experiment takes the following form:

$$y = (1 - 2x_{l,1} + 0.05\sin(4\pi x_{l,2}) - x_{s,1})^2 + (x_{l,3} - 0.5\sin(2\pi x_{s,2}))^2 + (x_{l,4} - 0.5\sin(2\pi x_{s,3}))^2 + x_{s,4} \qquad (16)$$

In each example, eight samples are generated using LHD for the $Input_s$s, and 40 samples of $Input_l$s are generated for each sample of $Input_s$s. The testing sample sizes of $Input_s$ and $Input_l$ are 40 and 40, respectively. The RMSEs of HL-SVR and SVR are shown Fig. 4. It is observed that the mean RMSE of HL-SVR is lower than that of SVR. The standard deviation of the HL-SVR is as well lower than SVR. The HL-SVR prediction approach shows better prediction accuracy and robustness.

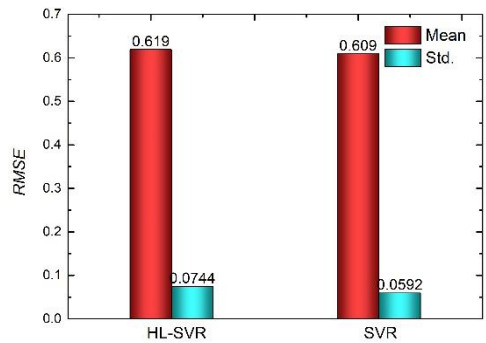

Figure 4 RMSE comparison between HL-SVR and SVR

## 4. Real data analysis (Dental implant)

4.1 Problem description

In this section, a computer simulation case, stress analysis of dental implant is used to validate the performance of HL-SVR. Dental implant is an important branch of dentistry that focuses on oral and maxillofacial prosthesis, and the stress at implant-bone interface is the key factor affecting its success ratio of implantation. Researchers used finite element method (FEM) to calculate and analyse the stress at implant-bone interface. However, since the high computational cost of FEM, the further analyses of dental implant based on FEM, such as optimization and global sensitivity analysis, are difficult to conduct. It is necessary to use other tools to replace

FEM, and SVR is a mature tool. For the stress at implant-bone interface, it is known that the structure and material of implant have important influence from the previous studies [16-18]. The structure parameters can be set as any value. However, the materials of dental implant can only be selected from several Ti alloys, so the material parameters of dental implant are limited [26-29]. Thus, the sample size of structure parameters is generally unequal to that of material parameters for dental implant. In this paper, HL-SVR is used to predict the stress at implant-bone interface based on the structural and material parameters of implant with unequal sample sizes, and the prediction results are compared with SVR to show its effectiveness and advances.

4.2 Finite element modeling

The mandible segment with dental implant is modelled as follows. The computerized image is created based the CT scanning data. The dental implant studied here is a two-hollow-cylinder dental implant composed of implant, abutment and screw as shown in Fig. 5. The crown is admitted considering that it is different for each tooth of each patient and many publications have reported the results of loading on the crown and that on the abutment are similar. The initial structure parameters of implant are as follows: the diameter and implanting length of dental implant are 4 mm and 8 mm, respectively; the length and the pitch of the thread on the implant are 6 mm and 1.2 mm; the cross-section of the thread is a regular triangle with side length of 0.6 mm.

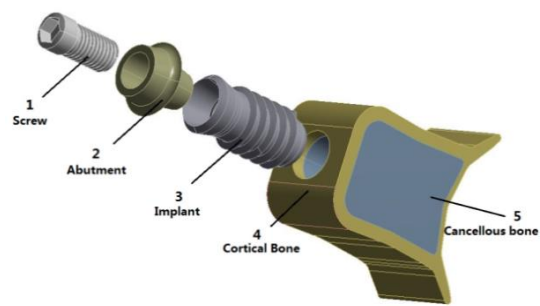

Figure 5 3D model for dental implant and mandible segment

The cancellous bone and cortical bone are regarded as anisotropic material and the parameters are listed in Table 4. The implant is assumed to be completely integrated with the bone, so the contacts among cancellous bone, cortical bone and implant are defined as bonded. The contacts among implant, abutment and screw are defined as fictional with friction coefficient of 0.3. The bottom surface of the mandible is fixed support, and the flanks of mandible at the direction of the mandible are fixed as frictionless support. 150N of occlusal force is loading on the upper surface of the abutment. After meshing the model, the thread contacts between bone and implant are refined further until the stress results are meshing independent. The index of stress at implant-bone interface is von-Mises stress (Fig. 6). From this figure, it can be found that the stress concentrations on the thread surface, especially the first thread facing the buccolingual direction, and decreases along with the implanting direction. The computational time of one-time FEM is about 37 mins (Inter Core i7-6700, RAM 16GB). It is too high for the optimization and analysis of stress at implant-bone interface which requires massive iterative calculations. In this paper, the FEM is replaced by HL-SVR and SVR, and the maximum stress at implant-bone interface is set as the prediction objective.

Table 4 Material properties of jawbone

|  | Elastic modulus(GPa) | | | Shear modulus(GPa) | | | Poisson's ratio | | |
| --- | --- | --- | --- | --- | --- | --- | --- | --- | --- |
|  | $E_x$ | $E_y$ | $E_z$ | $G_{xy}$ | $G_{yz}$ | $G_{xz}$ | $V_{xy}$ | $V_{yz}$ | $V_{xz}$ |
| Cortical Bone | 17.900 | 12.700 | 22.800 | 5.000 | 5.50 | 7.40 | 0.180 | 0.310 | 0.280 |
| Canellous Bone | 1.148 | 0.210 | 1.148 | 0.068 | 0.068 | 0.434 | 0.055 | 0.055 | 0.322 |

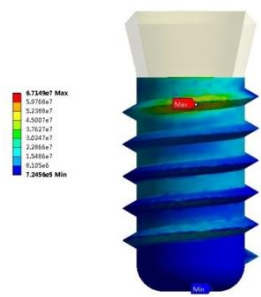

Figure 6 Stress at implant-bone interface

4.3 Design of experiments

According to the previous studies [26-33], the structure parameters of implant considered in this paper are as follows: the implanting length ($L_C$), the length of thread ($L_T$) and the pitch of thread ($P$), the bottom width ($L$) and the base angle ($β$) of isosceles triangle tooth. For material parameters of dental, the elastic modulus and the Poisson's ratio are considered. Eight materials are collected from previous studies as shown in Table 5. Two materials are selected randomly from Zr, TC4, Ti-Au, Ti (Grade 4), Ti-15Zr and Ti Alloy as the testing materials, and the others are used as training materials. To thoroughly test the performance of the proposed approach, all possible combinations of training materials and testing materials are considered. Thus, totally 15 experiments are designed (their details and experiment numbers are shown in Appendix A). In each experiment, LHS is used to generate the 25 training samples of structure parameters for each material (their design ranges are shown in Table 6). Thus, for each experiment, the training samples are 150, and the testing samples are 50.

Table 5 Material parameters of implant

| No | Material | Elastic modulus (GPa) | Poisson's ratio |
|---|---|---|---|
| 1 | Y-TZP [26] | 210 | 0.230 |
| 2 | Zr [27] | 200 | 0.310 |
| 3 | Ti (TC4) [28] | 110 | 0.350 |
| 4 | Ti-Au [29] | 106 | 0.340 |
| 5 | Ti (Grade 4) [30] | 105 | 0.360 |
| 6 | Ti-15Zr [31] | 102 | 0.335 |
| 7 | Ti alloy[32] | 91 | 0.230 |
| 8 | Ti-Nb-Zr [33] | 71 | 0.320 |

Table 6 Design range of structure parameters

|  | $L_C$ (mm) | $L_T$ (mm) | $P$ (mm) | $L$ (mm) | $β$ (°) |
|---|---|---|---|---|---|
| Max. | 9.000 | 7.000 | 1.400 | 0.800 | 70 |
| Min. | 8.000 | 6.000 | 1.000 | 0.400 | 50 |

4.4 Results

Figure 7 shows the RMSE comparison between HL-SVR and SVR for the totally 15 experiments. From this figure, it can be seen that most RMSEs of HL-SVR and SVR are higher than 2.7, which indicates the effectiveness and feasibility that using them to replace FEM to estimate the stress at implant-bone interface. It is observed that the RMSEs of the HL-GPR approach are smaller than those of SVR for most experiments, but SVR outperforms HL-SVR for experiments NO.6~8. The mean and standard deviation are calculated and shown in Figure 7. Form this table, it can be seen that the mean RMSE of HL-SVR is much lower than that of SVR, but the standard deviation of HL-SVR is larger than SVR. The HL-SVR outperforms SVR in term of accuracy, but shows worse performance on the robustness of prediction accuracy. To statistically compare the performance between HL-SVR and SVR, Student's test and Wilcoxson test are both used in this paper. RMSE is used as the test index, and the null hypothesis is that SVR is worse than HL-SVR. The obtained P-values of Student's test and Wilcoxson test are 0.008479 and 0.004272, respectively, which are much smaller than 0.05. Thus, the null hypothesis, SVR is worse than HL-SVR, is accepted. In summary for the case examined, the HL-SVR presents a very promising tool to achieve excellent agreement with the computations from complex computer experiments (i.e., finite element modeling).

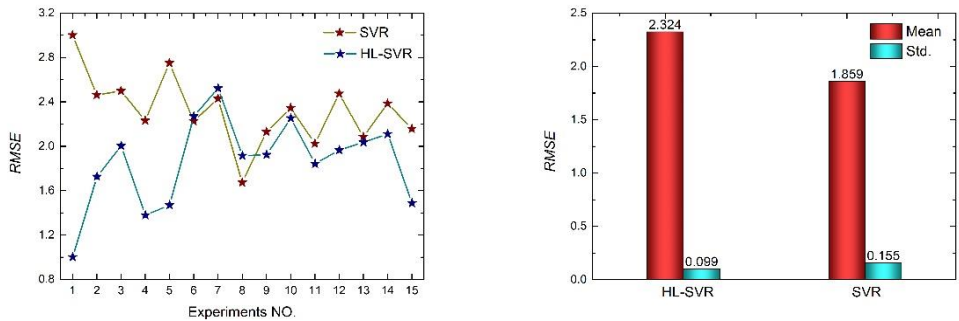

Figure 7 RMSE comparison between HL-SVR and SVR

Table 7 Results of Student's test and Wilcoxson test

|                | Null hypothesis         | Alternative hypothesis   | P-value  |
|----------------|-------------------------|--------------------------|----------|
| Student's test | SVR is worse than HL-SVR | SVR is better than HL-SVR | 8.479E-3 |
| Wilcoxson test | SVR is worse than HL-SVR | SVR is better than HL-SVR | 4.272E-3 |

## 5. Conclusions

In this paper, a new prediction approach based on support vector regression, namely as high-low-level support vector regression approach (HL-SVR) is proposed for improving the performance of SVR on the problem that the input parameters have unequal sample sizes. The proposed approach is consisted of several low-level SVR models for the input parameters of larger sample sizes and one high-level SVR model for the input parameters of smaller sample sizes. For each training point of the input parameters of smaller sample sizes, one low-level SVR model is built based on its corresponding training samples of input parameters of larger sample sizes and responses of interest. To obtain the response of a new point, its input parameters of larger sample sizes are first inputted into the low-level SVR models. The obtained responses and the training samples of the input parameters of smaller sample sizes are used to build the high-level SVR model. The response of the new point is obtained through the high-level SVR model. To validate the performance of HL-SVR, several numerical examples. The mean Root mean squared error was used to evaluate the accuracy of the proposed approach, and the standard deviation of Root mean squared error was used to evaluate the robustness. The results show that HL-SVR performs better than SVR for these test problems in terms of both accuracy and robustness for numerical examples. One computer simulation, stress analysis at implant-bone interface in dental implant, is used to validate the performance of HL-SVR in engineering applications, the results indicate that HL-SVR presents higher prediction accuracy than SVR in terms of prediction accuracy, Student's test and Wilcoxson test. The proposed approach can effectively improve the prediction of SVR on the issue that the input parameters have unequal sample sizes.

**Acknowledgments**

**Appendix A**

Table 1 DoE of dental implant

| No. | Training materials | Testing materials |
|-----|--------------------|-------------------|
| 1   | 1,4,5,6,7,8        | 2,3               |
| 2   | 1,3,5,6,7,8        | 2,4               |
| 3   | 1,3,4,6,7,8        | 2,5               |
| 4   | 1,3,4,5,7,8        | 2,6               |
| 5   | 1,3,4,5,6,8        | 2,7               |
| 6   | 1,2,5,6,7,8        | 3,4               |
| 7   | 1,2,4,6,7,8        | 3,5               |
| 8   | 1,2,4,5,7,8        | 3,6               |
| 9   | 1,2,4,5,6,8        | 3,7               |
| 10  | 1,2,3,6,7,8        | 4,5               |
| 11  | 1,2,3,5,7,8        | 4,6               |
| 12  | 1,2,3,5,6,8        | 4,7               |
| 13  | 1,2,3,4,7,8        | 5,6               |
| 14  | 1,2,3,4,6,8        | 5,7               |
| 15  | 1,2,3,4,5,8        | 6,7               |